\documentstyle[11pt,epsfig]{article}
\newcommand \be{\begin{equation}}
\newcommand \bea{\begin{eqnarray}}
\newcommand \ee{\end{equation}}
\newcommand \eea{\end{eqnarray}}
\newcommand{\lp}{\left(}
\newcommand{\rp}{\right)}

\begin{document}

\title{Probing human response times}

\author{Anders Johansen \\
Teglg\aa rdsvej 119, 3050 Humleb\ae k, Denmark}
\date{\today}
\maketitle

\begin{abstract}
In a recent preprint \cite{eck}, the temporal dynamics of an e-mail network
has been investigated by J.P. Eckmann, E. Moses and D. Sergi. Specifically,
the time period between an e-mail message and its reply were recorded. It will 
be shown here that their data agrees quantitatively with the frame work 
proposed to explain a recent experiment on the response of ``internauts'' to 
a news publication \cite{www2} despite differences in communication channels,
topics, time-scale and socio-economic characteristics of the two population. 
This suggest a generalized response time distribution $\sim t^{-1}$ for human
populations in the absence of deadlines with important implications for 
psychological and social studies as well the study of dynamical networks.
\end{abstract}

\section{Introduction}

There can be little doubt that the World-Wide-Web (WWW) and Internet e-mail 
provides two of the most efficient methods for retrieving and distributing 
information. As such, it carries an enormous potential with respect to sale 
and marketing of all kinds of products. A rather troublesome feature from a 
research perspective of many ``old-age'' communication channels, such as 
newspapers, radio and TV, interpersonal contacts and so forth, is that the 
diffusion of information through these channels is in general very difficult 
and/or time consuming to probe. This is not so with the WWW and e-mail. The 
fact that these communication channels are computer-based and access in 
principle unrestricted  provides a rather unique opportunity to study in real 
time how {\it fast} individuals respond to a new piece of information. It is 
generally believed that the rate of diffusion of information in a given 
population depends on a number of socio-economical factors such as education, 
cultural status, exposure to mass media and interpersonal channels 
etc. \cite{rogers}. The evidence presented here propose otherwise.

Along a complementary line of research the WWW has provide a similarly unique
opportunity to study fast evolving (sociological or computer) networks 
\cite{netbooks}. Most studies of the WWW have until now focused on the more
easily accessed statistical properties such as the connectivity of the WWW (the
distributions of outgoing and incoming links etc.) and a consensus that the
WWW is scale-free has been established. As the WWW or any other 
sociological network is constantly evolving any such investigation can only 
deliver a ``snap-shot'' picture of the network. The assumption is then that
the dynamical features of the network is reasonably stable over the time-scales
considered and that the results are representative over time. However, in order
to correctly describe networks between interacting humans one must estimate 
characteristics specific to the nature of the ``nodes'', {\it e.g.}, 
psychological traits of humans, as one may otherwise be mislead when 
generalizing from a ``snap-shot'' analysis. In fact, little is known about
the internaut population's response to some external or internal event. With
respect to commercial exploitation of the WWW it is obviously the response of 
the internaut population to some new piece of information which is of prime
interest and not the topology or connectivity of the underlying network itself.

The purpose of the present paper is to provide for a characterization of the 
psychological/sociological traits of interacting humans in absence of formal
deadlines. Using empirical data of response time distributions of internaut 
individuals/populations when exposed to a new piece of information, a 
surprisingly robust ``law'' for human response times will be put forward.

\section{Response to an Internet Interview: Experiment I } \label{uncutexp}
\begin{figure}[t]
\begin{center}
\epsfig{file=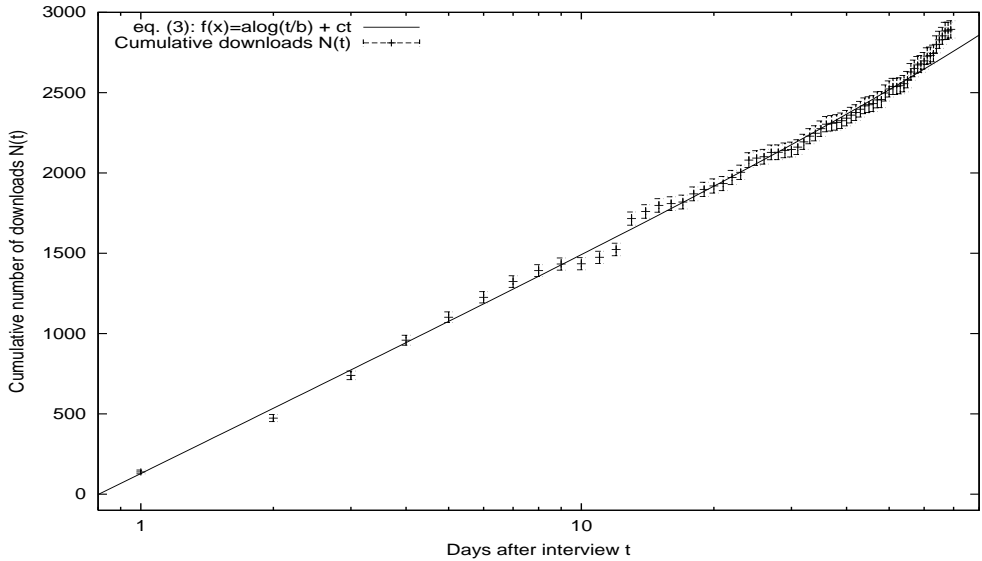,height=7.5cm,width=13.5cm}
\caption{\protect\label{uncutfig1} Cumulative number of downloads $N\lp t\rp$ 
as a function of time. The fit is $N\lp t\rp = a\ln\lp t/b \rp + kt$ with 
$a \approx 583$, $b \approx 0.80$ days and $k \approx 2.2$ days$^{-1}$.}
\end{center}
\end{figure}
The Nasdaq crash culminating on Friday the 14th of April 2000 caught many 
people with surprise and shook the stock market quite forcefully.  As always, 
many different reasons for the crash were given ranging from the anti-trust 
case against MicroSoft to an ``irrational exuberance'' of the participants
on the stock market. The author also had bid for the cause \cite{nascrash}, 
which was made public on Monday the 17th of April 2000 on the Los Alamos 
preprint server. As a result, a forty minute interview with the author called 
``The World (Not) According to GARCH'' was published on Friday the 26th of May 
2000 on a ``radio website'' (www.wallstreetuncut.com) In addition, the URL to 
the author's 
papers was announced making it clear that work on stock market crashes in 
general and the recent Nasdaq crash in particular could be found using the 
posted URL. As the main subject of the interview was a somewhat technical 
discussion about whether stock market crashes could be predicted, it seems 
reasonable to assume that the population in a socio-economic context was 
relatively homogeneous. The experiment proceeded as follows. The number of 
downloads of papers from the authors homepage as a function of time (days)
from the appearance of the interview was recorded. In figure \ref{uncutfig1}, 
the cumulative number of downloads $N\lp t\rp$  as a function of time $t$ after
the appearance of the interview is shown on a semi-logarithmic scale. 
The data is over one and half decade surprisingly well-captured by the relation
\be \label{downeq}
N\lp t\rp \sim a\ln\lp t/b \rp + ct \Rightarrow
n\lp t\rp = \frac{dN\lp t\rp}{dt} \sim \frac{1}{t} + c .
\ee
Here $k$ represents a constant background rate. After approximately $60$ days, 
the data breaks away from the fitted line. The reason is the appearance of the 
author's URL on the Social Science Research Network server (www.ssrn.com) 
causing a second advertisement. Hence, the experiment became influenced by an 
additional distribution channel and was consequently halted after 69 days.

\section{Response times in Email Interchange: Experiment II}

In a recent preprint \cite{eck}, the temporal dynamics of an e-mail network
has been investigated by J.P. Eckmann, E. Moses and D. Sergi. Specifically, 
the time period between an e-mail message and its reply was recorded. The
data set contains 3188 users interchanging 309129 messages and was obtained
from the log-files from one of the main mail servers of an university. In one
sense, this experiment is the ``cleanest'' as only a single distribution 
channel can exist. However, as each email message contains different 
information it is also the ``dirtiest'' with respect to the information 
distributed. Furthermore,  it seems reasonable to assume that the population 
in a socio-economic context is quite heterogeneous opposed to that of the first
experiment. In figure \ref{emailfig1}, we see that the cumulative distribution 
of time periods between an e-mail messages and its reply $N(t)$ can be modeled 
in terms of a response time $t$ of the recipient of the e-mail message,
\begin{equation} \label{cumu}
N(t) \sim a\log\left( \frac{ t + c}{b} \right) \Rightarrow
n\lp t\rp = \frac{dN\lp t\rp}{dt} \sim \frac{1}{t+c}.
\ee
Here the constant $c$ as a first approximation incorporates the fact that 
the measured time is not the true response time. The prime reason for the 
``shift'' $c$ is that most people do not download new e-mail messages 
instantaneously but instead every 10 minutes or so. Furthermore, some time is 
obviously needed in order to formulate and write the reply. The values of the 
fit parameters are $a=0.14$, $b=0.21$ and $c=0.25$, see figure caption for 
details of the fitting procedure. We clearly see that equation (\ref{cumu}) is 
an excellent approximation of the data over three decades. An 
additional feature of the data that can be rationalized is the first 
`` wiggle'' occurring around $10-16$ hours. Many people send e-mail messages 
just before leaving their work place. Since people generally share the same 
working hours (provided that they live in the same time zone), those messages 
are not answered before the next day. 
The main result to be extracted from this experiment compared to the previous
one is that the response time distribution of the population exposed to new 
information is {\it independent} of the specific nature of the new information.
This suggest that the observed behaviour indeed has its origin in psychological
traits of the population. Furthermore, it should be noted that whereas the 
previous experiment deals with time scales of up to 2 months, this experiment 
deals with time scales up to only one week, {\it i.e.}, a difference in
time scales of approximately a factor of ten. Nevertheless, the same 
distribution is obtained. This suggest that the difference in time scales are
mainly due to differences in communication channels, {\it i.e.}, impersonal or
personal.
\begin{figure}[t]
\begin{center}
\epsfig{file=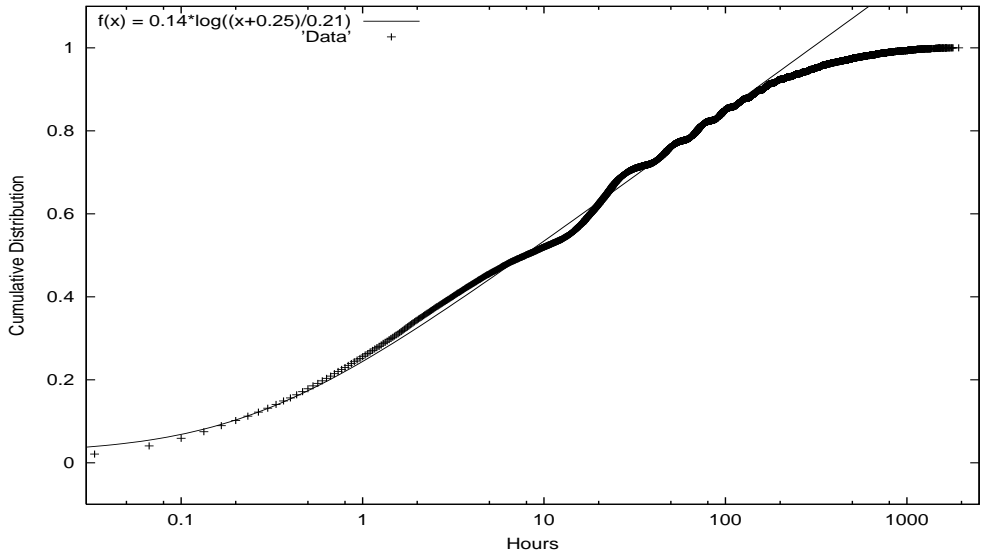,height=7.5cm,width=13.5cm}
\caption{\protect\label{emailfig1} Cumulative distribution of responses as a
function of time. The fit is $N\lp t\rp = a\ln \lp \lp t + c \rp / b 
\rp$ with  $a\approx 0.14$, $b\approx 0.21$ hours and $c \approx 0.25$ hours. 
Due to the ``wiggles'', the fit has been stabilized by first estimating $c$ 
from the data and then fitting $a$ and $b$ keeping $c$ fixed.}
\end{center}
\end{figure}

\section{Discussion}

The results shown in figures \ref{uncutfig1} and \ref{emailfig1} suggests that 
the general response rate of a human population $n\lp t\rp$ is that of a power 
law  $n\lp t\rp \sim 1/(t+c)$ where $c$ depends on the details of the 
distribution channel(s) used. Such power law dependence of the rate of 
``events'' are found in both natural and sociological systems. Two quite 
different examples are the Omori law for the rate of aftershocks as a function 
of time elapsed since the main shock 
\cite{omori} and the distribution of returns in the SP500 exceeding a specified
threshold as a function of time elapsed since the ``main event'', {\it i.e.}, 
the ``crash'', \cite{lilo}. A more appealing analogy, at least on a qualitative
level, is provided by the relaxation of spin glasses subjected to a magnetic 
field: At time $t$ after the appearance of the interview (experiment I), the 
exposed population consists of two groups, namely those who have not downloaded
a paper and those who have. Similarly with respect to experiment II, at any 
time $t$ the population considered consists of two groups, namely those who 
have an e-mail to answer and those who have not. The transition from the first 
state to the second demands the crossing of some threshold specific to each 
individual. We thus imagine that the announcement of the URL/the reception of 
emails plays the role a ``field'' to which the exposed population is subjected 
and study the relaxation process by monitoring the number of downloads/the 
number of replies as a function of time. Hence, we may view the process of 
downloading/replying as a diffusion process in a random potential, where the 
act of downloading/replying is similar to that of a barrier-crossing in the 
Trap model of spin glasses \cite{trap}. 
The most pressing unanswered question raised by the two experiments is whether 
the observed power law only is a characteristic of the entire population or if 
it is also true for an individual over time. A qualitative argument that 
``ensemble averaging'' may be the same as ``time averaging'' on the individual 
level (``ergodicity'') is that people react very differently to the same piece 
of information, {\it i.e.},  one cannot generally deduce {\it why} people react
on a specific piece of information. The reason is that individuals perceive the
same piece of information quite differently due to individual psychology 
despite similar socio-economical status, {\it e.g.}, not all rich people have 
an ambition to get richer because of the risk and/or work-load involved. This 
interpretations is supported by the results from the second experiment, where 
the new piece of information is changes with time and individual and 
{\it nevertheless} we get the same functional relationship between the time of 
the presentation of the new information and the response time distribution of 
the population. However, it seems {\it a priori} a quite formidable task to 
empirically verify whether these considerations are valid or not. 
\noindent

{\bf Acknowledgment:} I thank D. Sergi, E. Moses and J.P. Eckmann for the 
data of experiment II.

\end{document}